\documentclass[amsmath, amssymb, twocolumn, aps, pre]{revtex4-2}

\usepackage{graphicx}
\usepackage{dcolumn}
\usepackage{bm}
\usepackage{hyperref}
\usepackage{dsfont}

\usepackage{tikz}   
\usetikzlibrary{shapes}
\usepackage{tikz-cd}  
\usepackage{adjustbox} 
\usepackage{pgfplots}
\usepackage{pifont}
\newcommand{\cmark}{\ding{51}}%
\newcommand{\xmark}{\ding{55}}%

\DeclareRobustCommand\dotRed {\tikz \fill[red]             (0,0) circle (0.1);}
\DeclareRobustCommand\dotViolet {\tikz \fill[violet]             (0,0) circle (0.1);}

\DeclareRobustCommand\TriaDownGreen{%
\tikz \fill[black!50!green,scale=0.1] (-1,1) -- (1,1) -- (0,-1) -- (-1,1);}

\DeclareRobustCommand{\SquareYellow}{%
\tikz {\fill[color=black!20!yellow] (0,0) circle (0.03);
\node[thick, scale=0.7, regular polygon, regular polygon sides=4, draw, color=black!20!yellow] at (0,0) {};}}

\newcommand{\ie}{\textit{i.e.} }
\newcommand{\tr}{\mbox{tr}}

\newcommand{\abs}[1]{\left|#1\right|}
\newcommand{\avr}[1]{\left\langle#1\right\rangle}
\newcommand{\ket}[1]{|#1\rangle}
\newcommand{\bra}[1]{\langle #1|}
\newcommand{\braket}[2]{\langle #1| #2 \rangle}
\newcommand{\mc}[1]{\mathcal{#1}} 
\newcommand{\bs}[1]{\boldsymbol{#1}}
\newcommand{\T}{\mc{T}}
\newcommand{\K}{\mc{K}}
\newcommand{\LQ}{\Lambda_\mc{Q}}
\newcommand{\LT}{\Lambda_\mc{T}}
\newcommand{\one}{\mathds{1}}
\newcommand{\proj}[2]{{\vert #1 \rangle \langle #2 \vert}}

\begin{document}

\preprint{APS/123-QED}

\title{Symmetry operations and Critical Behaviour in Classical to Quantum Stochastic Processes}

\author{Gustavo Montes}
 \email{gustavo.montes@academicos.udg.mx}
 \affiliation{%
 Departamento de F\'\i sica, Universidad de Guadalajara, Guadalajara Jal\'\i sco, C.P.-44430, M\'exico
 }
\author{Soham Biswas}%
 \email{soham.biswas@academicos.udg.mx}
\affiliation{%
 Departamento de F\'\i sica, Universidad de Guadalajara, Guadalajara Jal\'\i sco, C.P.-44430, M\'exico
}%
\author{Thomas Gorin}%
 \email{thomas.g@academicos.udg.mx}
 \affiliation{%
 Departamento de F\'\i sica, Universidad de Guadalajara, Guadalajara Jal\'\i sco, C.P.-44430, M\'exico
}%

\begin{abstract}
Recently, a novel construction scheme for generating quantum analogs of classical stochastic processes has been introduced.
Here, we use this scheme in order to generate a large class of self-contained quantum extensions of a classical Markov chain process using symmetry operations. We show that the relaxation processes unfold very differently for the different quantum extensions. This is supported by monitoring the coherence, the probability of reaching the equilibrium, the decay of the number of domain walls and the purity. 
Unexpectedly, we find a rather ambiguous relation between the coherence measure based on the L1-norm and the speed of the relaxation process.
Finally we find that the finite size scaling of the coherence measure exists for both short and long times and the value of the critical exponent is different for the short and long time.

\end{abstract}

\maketitle


\section{\label{sec:introduction} Introduction
}

In this paper, we investigate the frontier between quantum and classical 
stochastic systems. On a fundamental level, stochasticity is not intrinsic to classical dynamics, it rather emerges due to the chaotic dynamics of many particle
systems. In quantum mechanics, by contrast, genuine stochasticity appears as soon as measurement processes are included (wave function collapse). Although the measurement process can be avoided in many different ways, for instance by  including the measurement apparatus into the quantum system to be described, at some point, it has to pop up, if meaningful experimental results are to be discussed \cite{Schlosshauer05,Leggett05,Bassi13}.

A gradual transition from quantum dynamics (deterministic and unitary) to 
(partially) stochastic processes is usually the subject of study in the theory of 
open quantum systems. In this case, information about the dynamics of the quantum
system is leaking out into the environment, a process which may be seen as applying
measurements continuously with low probabilities. The characterization of such Markovian 
processes has been established in Refs.~\cite{SuMaJa61,GoKoSu76,Li76}, and is by now, well established~\cite{Breuer02} and used in countless applications.

Recently, an alternative approach (to describe systems intermediate between
quantum unitary and classical stochastic) has been proposed \cite{MontesBiswasGorin22}. 
There, one starts from the classical stochastic process, and asks for the 
(partially) quantum processes (quantum extensions) in its vicinity. 
This approach leads to processes which are close to classical but 
still quantum. In other words, they operate with finite, but very low amounts of 
coherence, as measured by the L1-norm based measure introduced in \cite{BaCrPl14}). 
Surprisingly, however, these processes show macroscopic behavior that is
different (even the scaling with system size) as compared to the classical mother process. 

The finding of stochastic processes which are close to classical (in terms of
the coherence measure) and still show very different properties on the
macroscopic scale (thermodynamic limit) is important for many reasons : (i) biochemical processes in warm and hostile environments are capable of producing
sufficient coherence to make transport processes efficient
(light harvesting, etc.)
\cite{engel07, cao20}
(ii) Superior quantum computation with low-coherence quantum computers, to name
two \cite{Preskill18}.

In the present contribution, we construct a large self-contained set of quantum
extensions of the classical relaxation process in the Ising model. These
extensions are generated from two single qubit operations, the Hadamard and the
NOT gate. We then investigate the resulting processes by measuring several
quantities as a function of time: (i) the probability to find the system in the
ground state; (ii) the decay of the number of domains; (iii) the purity; and
(iv) the coherence. We find that the amount of coherence is related to the degree
to which the quantum extension deviates from the purely classical process, and we
show the finite size scaling for its decay. 

Section~\ref{sec:Markov stochastic processes}, briefly introduces the concept of
classical Markov processes. Subsequently, the
concept of a quantum Markovian process and its connection with the formalism of open
quantum systems is introduced. Finally, a brief explanation of the
concept of quantum extension is given, which allows us to study quantum versions
in the vicinity of a classical process.  In Section III, the method of quantum
extensions is applied to an Ising chain subject to a zero-temperature quench,
and a strategy for constructing multiple quantum extensions is shown. In the first
part of section IV, exact numerical simulations are performed for chains of length
$N=12$, and the behavior of several macroscopic observables is analyzed.
Subsequently, the scaling behavior of coherence is investigated for spin chains up to
length $N=20$. In Section V, conclusions are presented and some
perspectives have been discussed.


\section{\label{sec:Markov stochastic processes} Classical and quantum stochastic processes}

Following~\cite{MontesBiswasGorin22}, we consider classical stochastic processes with a finite sample space, 
$\{\ket{j}\}_{j=1, \ldots ,d}$, where each $\ket{j}$ represents a possible configuration of the physical system. In this case, the dynamics of the system can be characterized by a 
sequence of stochastic transition matrices, $\{ \mathcal{T}(n)\, |\, n\in\mathbb{N}_0 \}$, such that
$\bs{p}(n+1) = \mathcal{T}(n)\; \bs{p}(n)$, i.e.
\begin{equation}
\forall\; 1\le i\le d\quad :\quad p_i(n+1) = \sum_{j=1}^d \mathcal{T}_{ij}(n)\; p_j(n)\; .
\end{equation}
Here, the component, $p_j(n)$, of the vector $\bs{p}$ denotes the probability that the system is found
in the configuration $\ket{j}$ at the discrete time $n$. 
The matrix element $\mathcal{T}_{ij}(n)$ is the conditional probability that the systems will be found in
the configuration $\ket{i}$ at time $n+1$, provided it is in configuration $\ket{j}$ at time $n$. That
means that the column vectors of any transition matrix have non-negative entries which sum up to one.
This guaranties the conservation of probability along the process.

The stochastic process can be generated from the composition of subsequent transition matrices, realized
by matrix multiplication~\cite{Kampen92}. That is, with $n \ge m \ge 0$:
\begin{align}
\label{Eq:CSProcess}
    \bs{p}(n+1) = \mathcal{T}(n)\; \mathcal{T}(n-1)\; \cdots\; \mathcal{T}(m)\; \bs{p}(m)\; .
\end{align}
Such a process is also called a ``Markov chain''.

\subsection{Quantum processes}

In quantum mechanics, the analog of a Markov chain is conveniently described within the framework of 
quantum channels~\cite{Breuer02,bengtsson2017, heinosaari2011}. Here, the probability vectors are 
replaced by density matrices $\varrho\in\mc{S}(\mc{H})$~\cite{vonNeumann2018,sudarshan1961}, where $\mc{H}$
is the Hilbert space of all linear combinations of classical configurations $\{\ket{j}\}_{j=1,\ldots,d}$, equipped with 
the scalar product where $\braket{i}{j}=\delta_{ij}$. 

The space $\mc{S}(\mc{H})$ is made up of all convex combinations of one-dimensional projectors in $\mc{H}$. That is
\begin{equation}
    \mc{S}(\mc{H}) = \Big \{ \varrho = {\textstyle \sum_j}\, p_j\; \frac{\ket{\psi_j}\bra{\psi_j}}{\|\psi_j\|^2}\; \Big |\; 
    {\textstyle \sum_j}\, p_j = 1\, ,\; \psi_j\in\mc{H}\, \Big \}\; .
\end{equation}
Equivalently, we may say that $\varrho\in\mc{S}(\mc{H})$ if and only if $\varrho$ is a positive, Hermitian
operator on $\mc{H}$, with unit trace. A quantum Markov chain can then be constructed from quantum 
channels, i.e. completely positive and trace preserving (CPTP) linear maps on $\mc{S}(\mc{H})$~\cite{davies1969}.
Hence, instead of evolving probability vectors, we are now evolving density matrices in time, such that 

\begin{align}
    \label{Eq:Quantum-process-def}
    \varrho(n+1) &= \LQ(n)\circ\cdots\circ\LQ(m)\, [\varrho(m)]\; ,
\end{align}
where $\circ$ denotes composition~\cite{bengtsson2017, heinosaari2011} and the discrete times
are ordered as $n\le m\le 0$.

In what follows, we will make use of the fact that 
every CPTP map admits a sum representation in terms of Kraus operators \cite{kraus1983}. Then, for each element 
in the Markov chain given in Eq.(\ref{Eq:Quantum-process-def}), we can write
\begin{align}
    \LQ(n)[\varrho(n)] &= \sum_\alpha \K_\alpha(n)\,\varrho(n)\,\K_\alpha^\dagger(n)\,, 
\end{align}
where the Kraus operators, $\{\K_\alpha(n)\}$, satisfy the condition 
\begin{align}\label{Eq:CPTP-condition-for-KrausOperators}
\forall\; n\in\mathbb{N}_0\quad :\quad 
\sum_\alpha \K_\alpha^\dagger(n)\K_\alpha(n) = \one\; . 
\end{align}

Note that the quantum Markov chains defined here, include classical processes such as the
ones defined in~Eq.(\ref{Eq:CSProcess}). To see this, we first express the probability vector, 
$\bs{p}$, as a diagonal density matrix, $\bar\varrho$, and subsequently replace the transition matrices, 
$\T(n)$, by the quantum channels $\LT(n)$, 
\begin{align}
    \label{classicalMap-KR}
    \bar\varrho(n+1) = \Lambda_{\T}(n)[\bar\varrho(n)] = \sum_{i,j} \K_{ij}(n) \bar\varrho(n) 
    \K_{ij}^\dagger(n),
\end{align}
with the following definition for the Kraus operators
\begin{align}
    \label{classicalMap-KrausElement}
    \Big\{ \K_{ij}(n) &= \sqrt{\T_{ij}(n)}\proj{i}{j} \Big\}_{1\le i,j\le d}\, , 
\end{align}
where it is straight forward to show that condition 
(\ref{Eq:CPTP-condition-for-KrausOperators})
is always satisfied. Note that the maps $\LT(n)$ belong to the set of incoherent CPTP maps, which are neither been able to create nor detect coherence~\cite{BaCrPl14, winter2016}.

\subsection{Quantum extensions}

We are interested in quantum processes which reduce to a given classical process when observed or measured sufficiently often. We name such processes ``quantum extensions'' of the given
classical stochastic process. The idea is that if we observe a classical stochastic process, we may
think that its stochastic nature really comes from a quantum process, which has lost its coherence
due to direct or indirect (i.e. coupling to some environment) measurements. The quantum extensions we are interested in, may then be seen as those quantum processes, where measurements are not 
sufficiently complete or frequent, such that the resulting process is only partially incoherent.
This idea can be clarified by the following formal definition~\cite{MontesBiswasGorin22} : 

\paragraph*{Definition:} A quantum process described by a sequence of quantum
maps, $\Lambda_Q(n)$ is a quantum extension of the classical process described by $\Lambda_{\T}(n)$ if
and only if
\begin{align}
    \label{Def:QuantumExtension}
    \forall\,n \quad : \quad \Lambda_{\T}(n) = \mc{P}\circ\Lambda_Q(n)\circ\mc{P}\,,
\end{align}
where $\mc{P}$ denotes a complete measurement of the set of classical configurations (\ie, the basis 
states $\ket{j}$).\\ 

To find quantum extensions for a given classical Markov chain, we use the following guiding principles. 
(i) We try to find quantum extensions for each stochastic map $\LT(n)$, individually. 
(ii) For the corresponding quantum maps, we try to replace as far as possible random “which path” decisions,
encoded in the matrix elements of $\T_{ij}(n)$, by superpositions of all available options. 
(iii) In order to obtain a valid quantum extension, we make sure that Eq.~(\ref{Def:QuantumExtension}) 
is fulfilled. 

As explained in \cite{MontesBiswasGorin22} it is not an easy task to verify this condition. To shed some light on this task, in the next section we analyze the relationship that must exist between the
Kraus operators of the quantum map $\LQ$ and the matrix elements of the classical transition matrix $\T$.

\section{\label{I} Implementation in the Ising model}

Consider a linear chain made up $N$ subsystems of two classical states, or spins. We assume the chain is embedded in equilibrium with a thermal bath at very high temperature ($T \rightarrow\infty$ ). In this condition, the chain is found in a disordered phase, i.e., each spin can be in one of its two possible classical states with equal probabilities. Then the system is quenched to zero temperature ($T = 0$) and as a result the chain enters a relaxation process to reach equilibrium. We model this relaxation process using the Glauber dynamics \cite{glauber} that have been widely used to study the zero-temperature dynamics for classical spin systems \cite{glexample, bnnni,Newman2000}. The process consists of the successive application a global stochastic map $\T(n) = \T_{\rm all}$, independent of $n$, which is a uniform 
mixture of local stochastic maps, $\T^{(q)}$, to be explained below. In other words,
\begin{equation}
    \T_{\rm all} = \frac{1}{N}\sum_{q=1}^N \T^{(q)}\; . 
\label{allunfform:cl}\end{equation}
This construction describes the procedure where one selects at random one of the spins in the chain,
and then updates the spin, according to Glauber's acceptance criterion. In this scenario, one Monte Carlo
time step (MCS) corresponds to $N$ applications of $\T_{\rm all}$, which in turn corresponds to $N$ local updates (including random selection) in the Glauber algorithm.

To define the local update operation, we enumerate the spins from $q=1$ to $N$, denote a classical
configuration of the spin chain by $\ket{\vec s} = \ket{s_1, s_2, \ldots, s_N}$ and adopt periodic boundary condition, such that $s_0 = s_N$ and $s_{N+1} = s_1$. The two possible states of each spin are $s_j\in\{ 0,1\}$, where 0 (1) denotes the spin pointing upward (downward). The local operation $\T^{(q)}$ involves the spin $q$
and its immediate neighbors. In the truncated configurational basis $\{ \ket{s_{q-1}, s_q, s_{q+1}} = \ket{000}, \ket{001},
\ket{010}, \ket{011}, \ket{100}, \ket{101}, \ket{110}, \ket{111}\}$, it reads
\begin{align}
	\label{TransitionMatrix3S}
	\T^{(q)} &=
\begin{pmatrix}
	1 &  0  & 1 &  0  &  0  & 0 &  0  & 0 \\
	0 & 1/2 & 0 & 1/2 &  0  & 0 &  0  & 0 \\
	0 &  0  & 0 &  0  &  0  & 0 &  0  & 0 \\
	0 & 1/2 & 0 & 1/2 &  0  & 0 &  0  & 0 \\
	0 &  0  & 0 &  0  & 1/2 & 0 & 1/2 & 0 \\
	0 &  0  & 0 &  0  &  0  & 0 &  0  & 0 \\
	0 &  0  & 0 &  0  & 1/2 & 0 & 1/2 & 0 \\
	0 &  0  & 0 &  0  &  0  & 1 &  0  & 1 
\end{pmatrix}\; .
\end{align}

To construct the quantum version(s) of the classical stochastic process, we 
translate Eq.~(\ref{allunfform:cl}) into the quantum channel setting, and search for 
quantum extensions of the local update operations $\T^{(q)}$. That is,
\begin{equation}
    \label{eq:QE-localUpdate}
    \LQ^{\rm all} = \frac{1}{N}\sum_{q=1}^N \LQ^{(q)}\; ,
\end{equation}
where the channels $\LQ^{(q)}$ should be constructed in such a way that 
random choices applied in case of the configurations, 
$\{\ket{001}, \ket{011}\}$ and $\{\ket{100}, 110\}$, are replaced by their respective 
superpositions. To this end, we assume that $\LQ^{(q)}$ can be described by a small
number of Kraus operators. Then we investigate whether Eq.~(\ref{Eq:CPTP-condition-for-KrausOperators}) (completeness of 
the Krauss decomposition) and Eq.~(\ref{Def:QuantumExtension}) (validity as a quantum extension) can be
fulfilled. 

In the case of only one Kraus operator, that operator must be a unitary. In the present case, this is
not possible because the classical map $\T_q$ maps two configurations $\ket{000}$ and $\ket{010}$ onto 
the same output configuration, $\ket{000}$. We will therefore concentrate on the case of two Kraus operators.
\begin{equation}
    \LQ^{(q)}[\varrho]=\K_1\, \varrho\, \K_1^\dagger + \K_2\, \varrho\, \K_2^\dagger \; .
\end{equation}
In that case, the most general form of the two operators is as follows:
\begin{subequations}
    \label{Eq:3SpinQuantumMap}
    \begin{align}
        \K_1 &= \begin{pmatrix}
            1  &   0    &  0  &   0    &   0    &  0  &   0    &  0  \\
            0  & X_{11} &  0  & X_{12} &   0    &  0  &   0    &  0  \\
            0  &   0    &  0  &   0    &   0    &  0  &   0    &  0  \\
            0  & X_{21} &  0  & X_{22} &   0    &  0  &   0    &  0  \\
            0  &   0    &  0  &   0    & X_{33} &  0  & X_{34} &  0  \\
            0  &   0    &  0  &   0    &   0    &  0  &   0    &  0  \\
            0  &   0    &  0  &   0    & X_{43} &  0  & X_{44} &  0  \\
            0  &   0    &  0  &   0    &   0    &  0  &   0    &  1  
        \end{pmatrix}\,,  \\
         \K_2 &= \begin{pmatrix}
            0  &   0    &  1  &   0    &   0    &  0  &   0    &  0  \\
            0  & X_{31} &  0  & X_{32} &   0    &  0  &   0    &  0  \\
            0  &   0    &  0  &   0    &   0    &  0  &   0    &  0  \\
            0  & X_{41} &  0  & X_{42} &   0    &  0  &   0    &  0  \\
            0  &   0    &  0  &   0    & X_{13} &  0  & X_{14} &  0  \\
            0  &   0    &  0  &   0    &   0    &  0  &   0    &  0  \\
            0  &   0    &  0  &   0    & X_{23} &  0  & X_{24} &  0  \\
            0  &   0    &  0  &   0    &   0    &  1  &   0    &  0  
        \end{pmatrix}\,.   
    \end{align}
\end{subequations}
Here we collect the potentially free parameters into a $4$$\times$$4$ matrix $X$.
We then find that in order to fulfill Eq.~(\ref{Eq:CPTP-condition-for-KrausOperators}) the column vectors of $X$ must be orto-normal, and in order to fulfill Eq.~(\ref{Def:QuantumExtension}), it must hold
\begin{equation}
    \forall\, 1\le i\le 2\, ,\; 1\le j\le 4\quad :\quad |X_{ij}|^2 + |X_{i+2,j}|^2 = 1/2\; .
\end{equation}
For the purpose of our analysis it is enough to consider all the elements of $X$ as reals.

\paragraph{Generating set of quantum extensions:}

As explained at the previous section, the core of the quantum construction is based
on two elementary operations. $\sigma_x$ that flip spins and  the Hadamard  gate $H$
that replace the statistical mixtures by superpositions. One can observe that 
combinations of these two operations, $\sigma_x H$ or $H\sigma_x$, are also valid that creates superpositions. Then, one wonders if it is possible to construct different quantum extensions based in these elemental operations. Considering the 
minimal set $g:=\bs{\{}\sigma_x, H\bs{\}}$ along with the multiplication operation, 
it is not difficult to show that the following group can be constructed
\begin{align}
\mc{G}:=&\bs{\{}\pm\one, \pm\sigma_x, \pm i\sigma_y, \pm\sigma_z, \pm H, \pm H\sigma_x, \pm \sigma_x H, \pm \sigma_x H \sigma_x\bs{\}}\,. 
\end{align}
You should notice that $\mc{G}$ contains the identity, $\one$, and given that 
every element of the set is unitary its inverse exists. 
To generate different quantum extensions in terms of the matrix $X$, here we 
propose that $X$ is an element of the set $\mc{G}^{\otimes 2}$, whenever the 
conditions to be a quantum extension defined in the previous section are satisfied.
The results are summarized in Table \ref{tab:Set-of-QE}. 
Take for example, the two basic quantum extensions defined in Ref. 
\cite{MontesBiswasGorin22}, 
HAD-0 and SYH-0 
which in our construction can be identified as $X\in\{H_0, S_0\}$ respectively,
where
\begin{align}
    \label{Eq:QE-H0}
        H_0 &= \one\otimes H = \frac{1}{\sqrt{2}}\begin{pmatrix}
        1 & 1 & 0 & 0 \\
        1 &-1 & 0 & 0 \\
        0 & 0 & 1 & 1 \\
        0 & 0 & 1 &-1 \\
    \end{pmatrix} \\
    \label{Eq:QE-S0}
        S_0 &= \one\otimes \sigma_x H = \frac{1}{\sqrt{2}}\begin{pmatrix}
        1 &-1 & 0 & 0 \\
        1 & 1 & 0 & 0 \\
        0 & 0 & 1 &-1 \\
        0 & 0 & 1 & 1 \\
    \end{pmatrix}\,.
\end{align}

\begin{center}
\begin{table}[]
    \caption{Quantum extensions generated by the combination of the positive elements in $\mc{G}$.
    The check-mark stands for the valid quantum extensions.}
    \label{tab:Set-of-QE}
    \begin{tabular}{l|cccccccc}
        $\bigotimes$ & $\one$ & $\sigma_x$ &$i\sigma_y$ &$\sigma_z$ & $H$ & $H\sigma_x$ & $\sigma_x H$ & $\sigma_x H \sigma_x$ \\
        \hline\hline
        $\one$      & \xmark & \xmark & \xmark &\xmark & \cmark &\cmark & \cmark & \cmark \\ 
        $\sigma_x$  & \xmark & \xmark & \xmark &\xmark & \cmark &\cmark & \cmark & \cmark \\ 
        $i\sigma_y$ & \xmark & \xmark & \xmark &\xmark & \cmark &\cmark & \cmark & \cmark \\ 
        $\sigma_z$  & \xmark & \xmark & \xmark &\xmark & \cmark &\cmark & \cmark & \cmark \\ 
        $H$         & \xmark & \xmark & \xmark &\xmark & \cmark &\cmark & \cmark & \cmark \\ 
        $H\sigma_x$ & \xmark & \xmark & \xmark &\xmark & \cmark &\cmark & \cmark & \cmark \\ 
        $\sigma_xH$ & \xmark & \xmark & \xmark &\xmark & \cmark &\cmark & \cmark & \cmark \\
        $\sigma_xH\sigma_x$ 
                    & \xmark & \xmark & \xmark &\xmark & \cmark &\cmark & \cmark & \cmark 
    \end{tabular}
\end{table}
\end{center}

\paragraph{Equivalent quantum extensions:} In the Table~\ref{tab:Set-of-QE}, 
we observe all the possible combinations that produces valid quantum extensions, 
but several of them are equivalent, such that we can reduce the set from 32 to 
the following set of 12 different quantum extensions
\begin{align}
    \label{eq:SetOfQE}
    \bs{\Big\{}\one&\otimes H, & \one&\otimes\sigma_xH \nonumber \\
    \sigma_x&\otimes H, & \sigma_x&\otimes\sigma_xH \nonumber \\
    i\sigma_y&\otimes H,& i\sigma_y&\otimes\sigma_xH \nonumber \\
    \sigma_z&\otimes H, & \sigma_z&\otimes\sigma_xH \nonumber \\
    H&\otimes H, & H&\otimes\sigma_xH \nonumber \\
    \sigma_xH&\otimes H,& \sigma_xH&\otimes\sigma_xH \bs{\Big\}}
\end{align}

At the beginning of the evolution all these quantum extensions introduce approximately the same amount of coherence into the system,
however, the behavior of the macroscopic observables for later times is very different, 
to the extent of being opposite with respect to the classical process.  
This apparent paradox is addressed by observing different dissipation rates in the coherences of the different quantum extensions. 

\section{\label{N} Numerical simulations} 
In the previous section we introduce a method that allows to generate several 
quantum extensions. Here we analyzed the exact evolution of the Ising chain under the influence of those quantum effects.

Considering a small chain length of $N=12$ spins and defining as initial 
ensemble, $\varrho_0$, the statistical mixture that containing only configurations with zero magnetization, the time evolution of the chain is obtained by the successive application of the uniform mixture of map $\LQ^{(q)}$, Eq.
(\ref{Eq:3SpinQuantumMap}), applied to all the sites along the chain as is
described in Eq.~\ref{eq:QE-localUpdate}.
As we will see below, the different quantum extensions modify the behavior of certain observables.  What is more important for us is to analyze the
possibility that quantum effects can accelerate or decelerate the relaxation process to the equilibrium. A phenomenon that has been previously reported
\cite{MontesBiswasGorin22}.

\begin{figure*}[ht]
    \centering
    \includegraphics[width=1.0\linewidth]{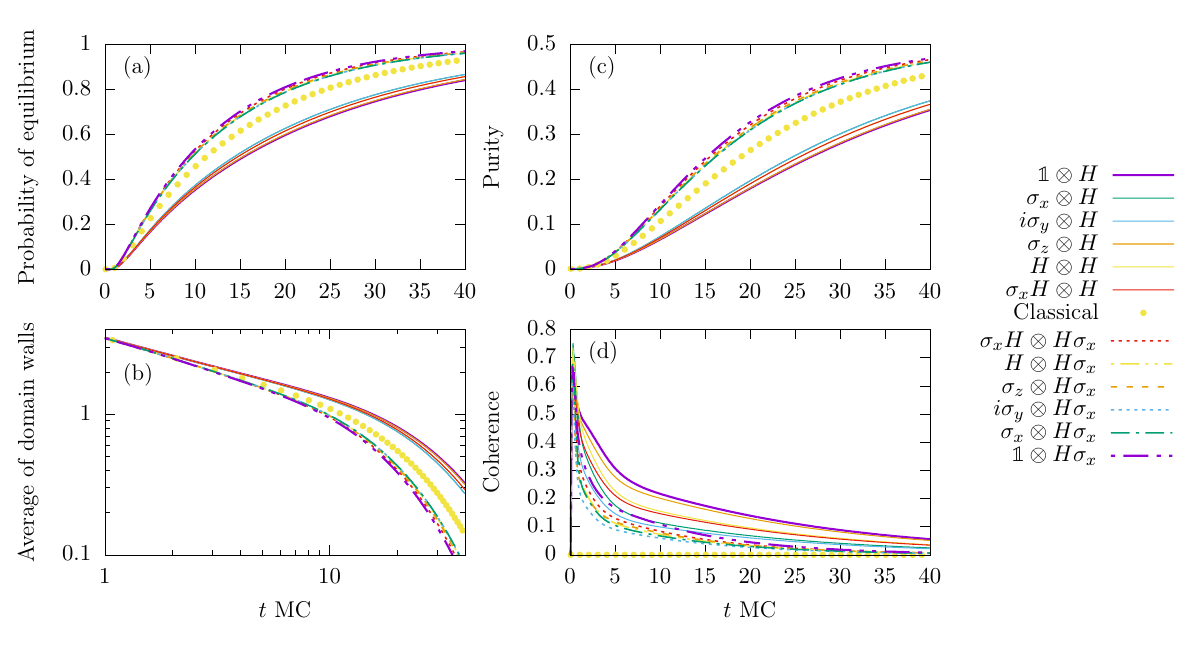}
    \caption{Observables measured for a linear chain of 12 spins.
    Panel (a), Probability to reach the equilibrium $vs\, t$ computed using
    Eq.~\ref{eq:ProbabilityOfEquilibrium}. Panel (b), Purity $vs\, t$ computed from 
    Eq.~\ref{eq:Purity}. Panel (c), Average of domain walls $vs\, t$ see 
    Eq.~\ref{eq:DomainWalls}. Panel (d), Coherence $vs\, t$ see Eq.~\ref{cohmeasure}.
    In all cases the time is measure in units of MC steps. The different curves corresponds
    to the different quantum extensions obtained in~\ref{eq:SetOfQE}.}
    \label{fig:Observables}
\end{figure*}

\paragraph{Probability to reach equilibrium.}
In the classical dynamics of the Ising chain, 
in the limit $t\rightarrow\infty$, the system evolves to one of the two steady 
states, $\ket{\bs{0}}\equiv\ket{00\ldots0}$ or $\ket{\bs{1}}\equiv\ket{11\ldots1}$.
In contrast in the quantum evolution the system finishes in a statistical
mixture of these two states. The probability to reach the equilibrium, 
$P_{\text{Eq}}(t)$ can be defined in terms of the density matrix $\varrho(t)$ 
and the projector to the equilibrium subspace $\hat{P}_{\text{Eq}}=\proj{\bs{0}}
{\bs{0}}+\proj{\bs{1}}{\bs{1}}$ as follows
\begin{align}
    \label{eq:ProbabilityOfEquilibrium}
    P_{\text{Eq}}(t) &= \tr\left[\hat{P}_{\text{Eq}}\cdot\varrho(t) \right]\,.
\end{align}
where, $\tr[A]$, stands for the trace of the operator $A$. Figure~\ref{fig:Observables}-a shows the behavior of the equilibrium probability as a function of time (in MC units). At the beginning of the evolution all curves, classical and quantum, behave similarly. After a few MC steps we see how the quantum curves move away from the classical one. Interestingly the quantum curves separate into 2 groups defined by the columns in Eq.~\ref{eq:SetOfQE} which differ only
by the operation $H$ or $\sigma_xH$ to the left of the tensor product symbol. In the first group which we identify with solid lines in Figure~\ref{fig:Observables}-a, it is
clearly observed that the continuous creation and annihilation of coherences slows dawn 
the relation process towards equilibrium. However, a clear acceleration of the
relaxation process is seen in the dashed curves. The extreme cases of deceleration and 
acceleration are defined by the quantum operations defined in Eq.~\ref{Eq:QE-H0} and 
Eq.~\ref{Eq:QE-S0} respectively. 

\paragraph{Decay of domain walls.}
Another observable that is widely used in classical studies of the relaxation process in
spin systems is the decay of the number of domain walls. A domain wall is understood as 
the interface between two spins pointing in opposite directions. Since at the beginning
of evolution we start in a disordered phase, the number of domain walls is maximal. 
As the spins begin to cluster into larger domains pointing in the same direction, the
number of walls decreases as, $t^{-1/z}$. The parameter $z$ is known as the domain growth exponent. In the quantum version, to analyze the decay of domain walls we define 
the following operator: $\hat{D}_W=\sum_j n_D(j)\proj{j}{j}$, where $n_D(j)$ is the 
number of domain walls in the $j$th configuration. Thus, we can calculate the 
expectation value of this observable as
\begin{align}
   \label{eq:DomainWalls}
   \avr{\hat{D}_W}(t) &= \tr\left[\hat{D}_W\cdot\varrho(t) \right]\,.   
\end{align}

The behavior of this observable is described in Figure \ref{fig:Observables}-b, for the
different quantum evolutions together with the classical one. For short times,
quantum versions coincide with the classical evolution, but for long times, we observe how the
quantum versions depart from the classical one in groups, such as in the previous case.

\paragraph{Purity.}
The next observable we analyze is the purity of the system, $\mc{P}$. The purity quantifies how mixed
the system is or how far the system is from being represented by a pure state. Defined as, 
\begin{align}
    \label{eq:Purity}
    \mc{P}(t) &= \tr\left[ \varrho^2(t) \right]
\end{align}
it takes the maximum value of 1 when the system is in a pure state and the minimum value of $1/d$, 
when the system is at the maximum statistical mixture, where $d=2^N$ is the dimension of the Hilbert
space. The maximum statistical mixture refers to a diagonal density matrix, where the system is found
with the same probability $1/d$ in any of the classical configurations. In classical dynamics, the
relaxation process sends the system to one of two possible equilibrium configurations, $\ket{\bs{0}}$
when all spins point up or $\ket{\bs{1}}$ when all spins point down. In quantum dynamics, the
relaxation process sends the system to a statistical mixture of these two configurations which we
denote as $\varrho(t\rightarrow\infty)=p\proj{\bs{0}}{\bs{0}}+(1-p)\proj{\bs{1}}{\bs{1}}$
(for $p\approx 1/2$). So for long times the purity saturates at the value 1/2, while its initial value
is close to the minimum value since we start with a statistical mixture of all configurations with zero
magnetization. The purity dynamics is shown in Figure \ref{fig:Observables}-c where the different
curves correspond to the different quantum extensions. The yellow dotted curve corresponds
to the classical dynamics, where followed by the local map $\LQ^{(q)}$, a projective measurement is
applied that destroys all possible coherence created in the previous step. The purity behavior is
consistent with the two previous cases. The same group of quantum extensions show an improvement in the relaxation process, while in the second group the relaxation process is even slower than in the
classical case. 

\paragraph{Coherence.}
To better understand the mechanism by which the relaxation process is modified we
study the coherence of the system, which is calculated in terms of the L1-norm
\begin{align}
    \mc{C}(t) &= \sum_{i\neq j} \abs{\varrho_{ij}(t)}\,,  
\label{cohmeasure}\end{align}
where $\varrho(t)$ is the density matrix that describes the state of the chain at time $t$.
Recently, it has been shown that coherence can be understood as a quantum resource that can
be used to perform quantum tasks \cite{BaCrPl14}, in analogy to how entanglement is the
necessary resource to implement quantum protocols, such as teleportation, coherence is the
necessary resource to modify the behavior of the relaxation process. 

In figure \ref{fig:Observables}-d we observe the dynamics of the coherence as a function
of time, in units of MC steps. At the beginning of the evolution, the local map $\LQ^{(q)}$
introduces coherence along the system through the action of operations defined in
\ref{eq:SetOfQE}. Then, the relaxation process starts to dominate the dynamics, and we
observe the decaying of the coherence after a couple of MC steps, with different decay rates
for the different quantum extensions.

One is tempted to think that the more coherence in the system, the faster the relaxation process
finds the state (subspace in our case) of equilibrium, as typically happens in quantum search
protocols \cite{Grover96,Bennett97,Jones98}. However, as we have seen so far, this is not always true.

\paragraph{ Scaling for the coherence measure}
We consider the scaling behavior for the measure of
coherence, defined in Eq.~(\ref{cohmeasure}) of the quantum extensions $S_0$ and 
$H_0$ for both short and long times. To study the finite size scaling we have simulated the Ising model for spin chains of length $N=12, ~14, ~16, ~18$ and 20. Here we use an 
unraveling method \cite{Breuer02, Molmer93}, to avoid the evolution of huge density 
matrices. Considering a chain of length $N$ we define as the basic unit of time 
a Monte-Carlo step (MC), which consist of the application of the map, 
$\LQ^{(q)}$, $N$-times. This is the minimum unit of time needed to observe relevant changes 
in the dynamics of the chain.

\begin{figure}
\includegraphics[width=0.48\textwidth]{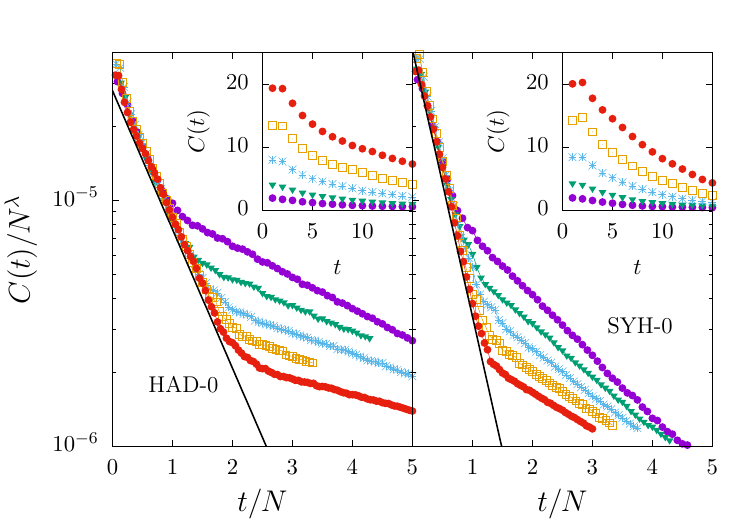}
\caption{Coherence measure as a function of time for different chain lengths, N = 12 ($\dotViolet$);
14 ($\TriaDownGreen$); 16 ({\color{cyan}$\ast$}); 18 ($\SquareYellow$); 20 ($\dotRed$). The inset 
shows the original data, for short times. The main plot shows the data collapse also at 
short times. }
\label{fig:CohST}\end{figure}

Figure \ref{fig:CohST} shows the measure of coherence [Eq. \ref{cohmeasure}] as 
a function of time. The scaled semi-log plot in the main figure, shows that the 
coherence decays exponentially for all times, with different decay rates at 
short and large times. The short time scaling behavior can be written as
\begin{equation}
 C(t) \sim N^{\lambda}\, \exp(- k\, t/N)
 \label{coST} \end{equation}
with $\lambda = 4.44 \pm 0.02$ for both $S_0$ and $H_0$ . However $k\simeq 2.5$ for $S_0$ and
$k\simeq 1.3$ for $H_0$. The main plot shows the collapse at the short time and inset shows 
the raw data.
 
\begin{figure}
\includegraphics[width=0.48\textwidth]{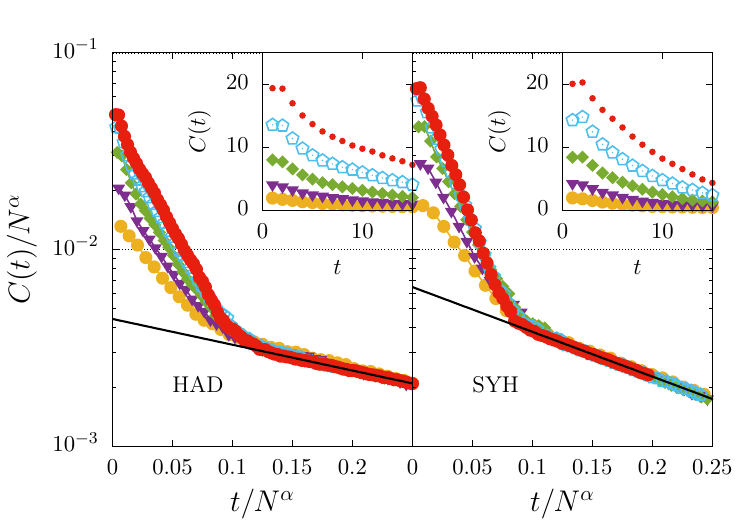}
\caption{Coherence measure as a function of time for different chain lengths, N = 12 ($\dotViolet$);
14 ($\TriaDownGreen$); 16 ({\color{cyan}$\ast$}); 18 ($\SquareYellow$); 20 ($\dotRed$). The 
inset shows the original data, for short times. The main plot shows the data collapse also at 
large times. }
\label{fig:Coherence-LargeTimes}\end{figure}

Late time scaling behavior for the coherence measure  have been studied for 
the first time in \cite{MontesBiswasGorin22}, see Fig. \ref{fig:Coherence-LargeTimes}, the 
results can be written as 

\begin{align}
    C(t) \sim N^\alpha \exp(-k_1 t/N^\alpha)
    \label{coLT}
\end{align}
with the following values for the parameters: $\alpha = 2.0 \pm 0.02$, $k_1 \simeq 3.02$ for 
$H_0$ and $\alpha = 1.91 \pm 0.014$, $k_1\simeq 5.23$ for $S_0$. 

Let us determine the crossover time $t_c(N)$. This is the time at which the decay of coherence 
cross from the initial behavior given by equation \ref{coST} to the long-time 
behavior given by equation \ref{coLT}. At the crossover time $t_c$, one can write 

\begin{equation}
 N^{\lambda}\, \exp(- k\, t_c/N) = N^\alpha \exp(-k_1 t_c/N^\alpha)
\end{equation}

By simplifying the above equation, one can write 
\begin{equation}
 t_c= (\lambda - \alpha)\dfrac{N}{k-k_1/N^{\alpha-1}}log(N)
\end{equation}

For both $S_0$ and $H_0$, $\alpha \sim 2$, for large $N$, neglecting the second term of the 
denominator, one can write
\begin{equation}
t_c \sim \frac{\lambda - \alpha}{k}\, N\, log(N)
\end{equation}
Note that the crossover time is a function of $N$, even for the large systems sizes, 
which indicates that behavior of coherence described by equation \ref{coST} will prevail 
for $N \rightarrow \infty$.

\section{\label{C} Conclusions} 

In this work, we constructed several different quantum extensions, derived from a real discrete subgroup of $U(2)$. These extensions include the two basic ones, $S_0$ and $H_0$, already discussed in Ref.~\cite{MontesBiswasGorin22}. 

We analyze the exact relaxation dynamics of the Ising model for $N=12$ spins, where we find that the behavior of characteristic macroscopic observables is different for the different quantum extensions. In particular, we study the time dependence of (i) the probability to find the system in an equilibrium state, (ii) the number of domain walls, (iii) the purity, and (iv) the coherence. In all cases, except for case (iv), the results are very similar within two antagonistic groups of quantum extensions. In the first group superpositions are generated with the gate $H$, in the second group with the gate $\sigma_x H$. By consequence the original extension $H_0$ belongs to the first group, where the relaxation process is slower than in the classical case and $S_0$ to the second, where it is faster. 

We include two quantum measures, the purity and the coherence (quantified as proposed in Ref.~\cite{BaCrPl14}). We expect the coherence to be particularly relevant for a quantum extension to show a different behavior than the original classical process. This is because it measures the distance between the density matrix with superpositions and the one where all superpositions (coherences) are removed. It is therefore natural to expect that the deviation from the classical behavior is directly related to the amount of coherence in the system. This expectation is confirmed only partially, as the behavior is rather heterogeneous among the different quantum extensions we have studied. 

In Appendix A, we analyze the evolution of the density matrix in the space of three collective coordinates. In this way, we hope to approach a possible explanation of the different behaviors of the quantum extensions. In fact, we found certain characteristic differences, in particular an unexpected symmetry in the $S_0$ case, absent for $H_0$. However, we are still far from understanding this phenomenon.

We then study the scaling behavior of the coherence for the two antagonist quantum extensions. We found that the finite size scaling form exists for both the short and long time. We not only observed different scaling behaviors for the short and the long time but also found the values of the critical exponent to be different in these two time regime. However, the values of the exponents are similar for both the quantum extensions $S_0$ and $H_0$. We calculated the crossover time which is a function of system size, even for large $N$.

\begin{acknowledgments}
The authors acknowledge the use of the Leo-Atrox super-computer at the CADS supercomputing 
center. This work has been financially supported by the Conahcyt project “Ciencias
de Frontera 2019,” No. 10872. G. M. thanks financial support from Conahcyt (Grant: 
"Estancias Posdoctorales por México 2022(1)").
\end{acknowledgments}

\bibliography{apssamp}

\appendix

\section{Exact evaluation of density matrix with respect to Hamming distances}

\begin{figure}
\includegraphics[width=0.48\textwidth]{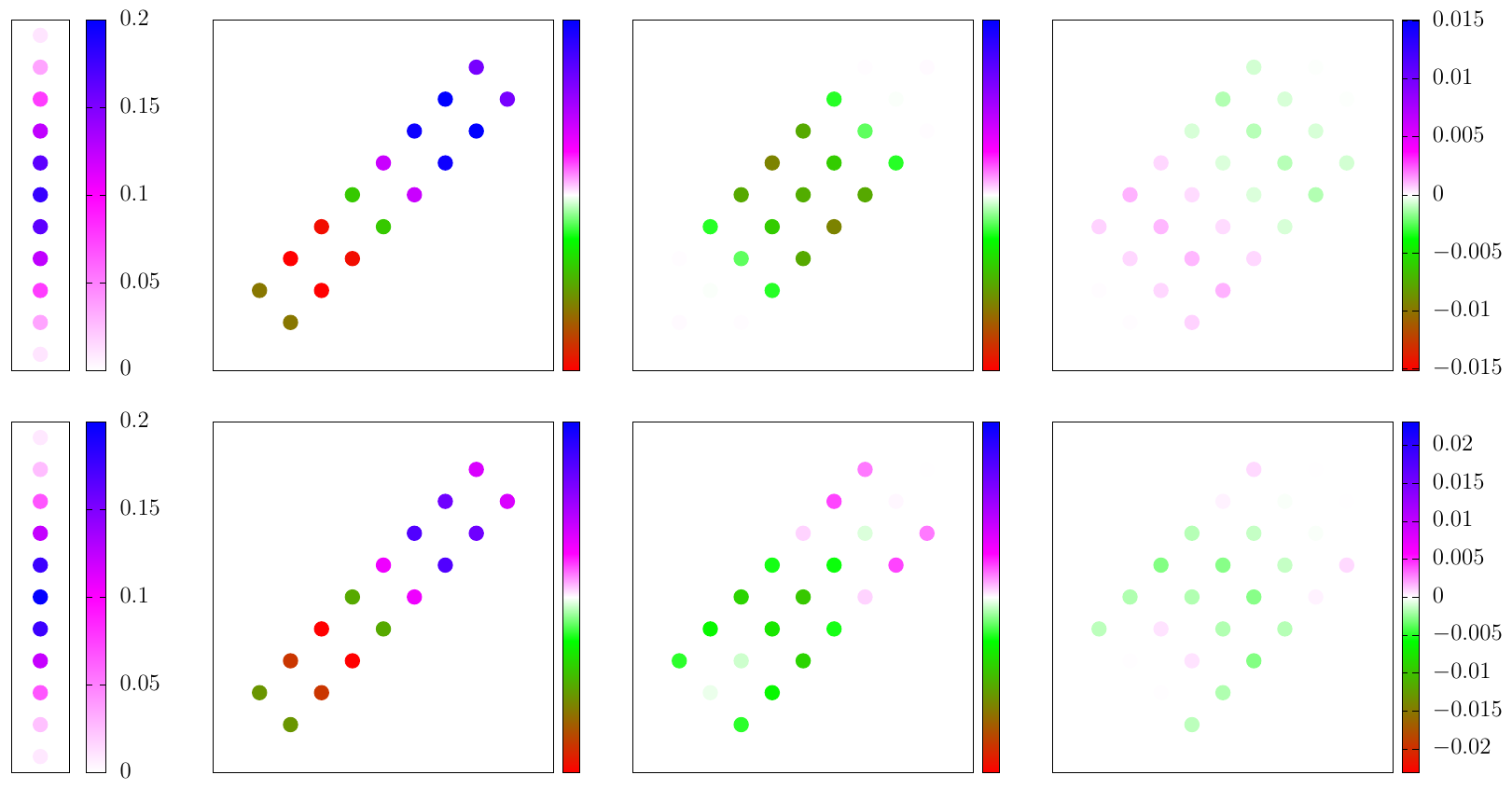}
\caption{Coherences after one MCS (Monte Carlo time step) for $N=10$ spins. $S_0$ (top row); 
$H_0$ (bottom row).}
\label{fig:Coh1MCS}\end{figure}

\begin{figure}
\includegraphics[width=0.48\textwidth]{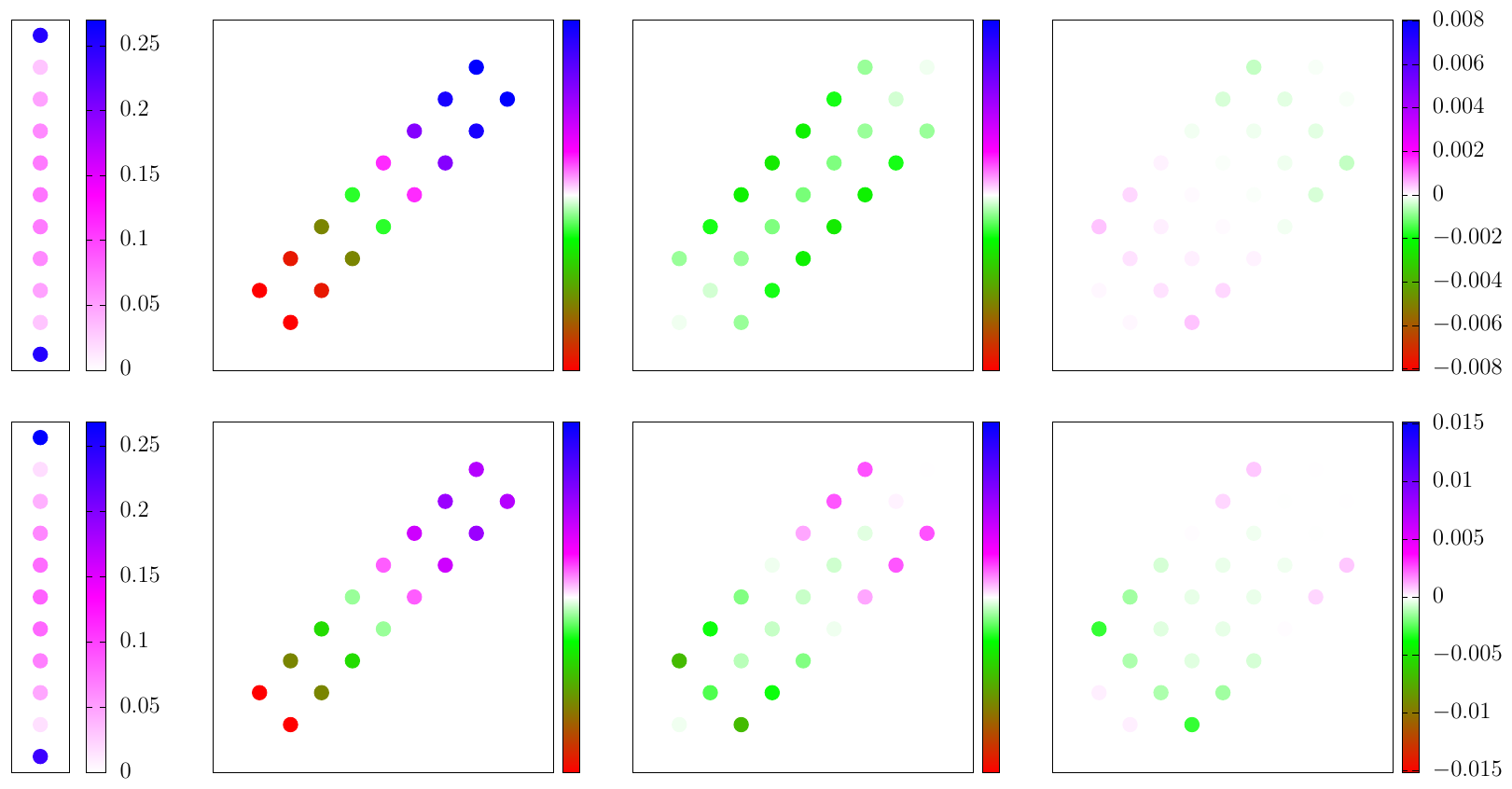}
\caption{Coherences after the respective half time for $N=10$ spins. $S_0$ (top row); $H_0$ (bottom row).}
\label{fig:CohThalf}\end{figure}

We consider the evolution of the system in terms of the density matrix
$\varrho(n)$ in the space of configurations (for $N$ spins, $2^N$
configurations). To generate the following figures, we classify all elements
$\varrho_{ij}(n)$ with respect to three Hamming distances: the distance
$d_{\rm H}(i,q_{\rm d})$ of the config. $i$ to the all-spins-down state
$i_{\rm d}$, the distance $d_{\rm H}(j,q_{\rm d})$, and the Hamming distance
$d_{\rm H}(i,j)$ between the configurations $i$ and $j$. Every class of matrix
elements is then characterized by the triple
$[ d_{\rm H}(i,q_{\rm d}), d_{\rm H}(j,q_{\rm d}), d_{\rm H}(i,j) ]$. The
figures~\ref{fig:Coh1MCS} and~\ref{fig:CohThalf} show that the average value of the matrix elements belonging to these classes.

For our simulations, the initial state is the uniform mixture of all
configurations with magnetization zero. The density matrix then only has
diagonal elements, and since all states have magnetization zero, their distance
to the all-spins-down state is $N/2$. Hence, in the space of the distances, these matrix elements correspond to the single point $(N/2,N/2,0)$.

After one elementary operation, from any of the zero-magnetization
configurations, one can arrive at a new state flipping a single spin. If the
operation is done with certainty, we ``move'' along the diagonal of the density
matrix towards the all-spins up or the all-spins down state. In this case, the
distances for $i$ and $j$ remain equal and either increase or decrease by one
unit. Alternatively, a unitary operation is applied, which produces non-diagonal
elements. Now the distance between $i$ and $j$ increases by one unit.

In the figures shown, the distances to the all-spins down state are mapped to the $x$- and $y$-axes, while the distance between $i$ and $j$ is held fixed: $d_{\rm H}(i,j) = 0$ (first column), $1$ (2nd column), $2$ (3rd column), and
$3$ (4th column).

In Figs.~\ref{fig:Coh1MCS} and~\ref{fig:CohThalf}, we compare the relaxation
dynamics when using $S_0$-gates (upper row) and $H_0$-gates (lower row) for $N=10$ spins.

In Fig.~\ref{fig:Coh1MCS}, we analyze the density matrix after one Monte Carlo
time step -- this means $N$ elemental operations. In this case, it is possible
that the system reaches one of the two minimum-energy configurations. However,
as one can see in the first column (diagonal elements), the system arrives at
those states $d_{\rm H}(i,j_0) = 0,N$ with very low probability, only. Here,
almost no difference can be observed between use of $S_0$-gates (upper row) or
$H_0$-gates (lower row). For the non-diagonal elements, the situation is different.
In all cases (column $2-4$) differences are observable, most clearly when
$d_{\rm H}(i,j) = 2,3$. Note that in the $S_0$-case, the pattern are strictly
symmetric (even distances) or anti-symmetric (odd distances) with respect to
the off-diagonal connecting the points $(0,N)$ (upper left corner) and $(N,0)$
(lower right corner). This symmetry is lost in the $H_0$-case. Note also the
slightly different scales for the average values of the matrix non-diagonal
elements which implies that the coherences are somewhat larger in the $H_0$ case.

Fig.~\ref{fig:CohThalf} shows the same quantities but at a later time,
$t_{1/2}$, when the probability to find the system in the minimum energy
subspace is equal to one half. For the $H_0$ case this is at a much later time than for the $S_0$ case. Again in the distribution for the diagonal
elements it is hard to observe any differences. But for the non-diagonal
elements, the amount of coherence is much larger for the $H_0$ case than the $S_0$ case (almost twice as much). We also observe that the symmetry in the pattern for the non-diagonal elements remains (is absent) in the $S_0$ ($H_0$) case.

\end{document}